\documentstyle[twocolumn,epsfig,fleqn,amstex,rotating,multirow]{mn}
\def\lsim{~\rlap{\raise 0.4ex\hbox{$<$}}{\lower 0.7ex\hbox{$\sim$}}~}
\def\gsim{~\rlap{\raise 0.4ex\hbox{$>$}}{\lower 0.7ex\hbox{$\sim$}}~}
\def\dd{{\rm d}}

\def\l0{L_\ast(0)}

\def\s0{S_\ast(0)}

\def\omg0{\Omega_0}

\def\a2{\alpha^{(2)}}

\def\mpc3{\ {\rm Mpc^{-3}}}
\def\gpc3{\ {\rm Gpc^{-3}}}

\def\lsim{~\rlap{\raise 0.4ex\hbox{$<$}}{\lower 0.7ex\hbox{$\sim$}}~}
\def\gsim{~\rlap{\raise 0.4ex\hbox{$>$}}{\lower 0.7ex\hbox{$\sim$}}~}
\def\dd{{\rm d}}

\def\l0{L_\ast(0)}
\def\hd{h_{_{\rm d}}}
\def\rs{r_{_{\rm s}}}

\def\s0{S_\ast(0)}

\def\omg0{\Omega_0}

\def\a2{\alpha^{(2)}}

\def\mpc3{\ {\rm Mpc^{-3}}}
\def\gpc3{\ {\rm Gpc^{-3}}}

\newcommand{\beq}{\begin{equation}}
\newcommand{\eeq}{\end{equation}}
\newcommand{\beqa}{\begin{eqnarray}}
\newcommand{\eeqa}{\end{eqnarray}}
\def\lsim{\lesssim}
\def\gsim{\gtrsim}

\begin{document}

\title[]
{ Searching for Neutrinos from WIMP Annihilations in the Galactic Stellar Disk}
\author[Myers {\it \& Nusser}]{Zacharia Myers and  Adi Nusser\\\\
Physics Department, Technion, Haifa 32000, Israel
and the Asher Space Research Institute\\
}
\maketitle

\begin{abstract}

Weakly interacting massive particles (WIMPs) are a viable candidate for the relic abundance of dark matter (DM) produced in the early universe. So far WIMPs have eluded direct detection through interactions with baryonic matter. Neutrino emission  from accumulated WIMP annihilations in the solar core has been proposed as a signature of DM, but has not yet been detected. These null results may be due to small scale DM density fluctuations in the halo with the density of 
our  local region being lower than the  average  (${\rm \sim 0.3 \rm\ GeV/cm^3}$). 
However, the  accumulated  neutrino signal from WIMP annihilations in the Galactic stellar disk would be insensitive to  local density variations. Inside the disk, dark matter can be captured by stars causing an enhanced annihilation rate and therefore a potentially higher neutrino flux than what would be observed from elsewhere in the halo. We estimate a neutrino flux from the WIMP annihilations in the stellar disk to be enhanced by more than an order of magnitude compared to the neutrino fluxes from the halo. We offer a conservative estimate for this enhanced flux, based on  the WIMP-nucleon cross-sections obtained from direct-detection  experiments by assuming a density of  ${\rm \sim 0.3 \rm\ GeV/cm^3}$ for the local DM. We also compare the detectability of these fluxes with a signal of diffuse high energy neutrinos produced in the Milky Way by the interaction of cosmic rays (CRs) with the interstellar medium (ISM). These comparative signals should be observable by large neutrino detectors.

\end{abstract}


\section {Introduction}
\label{sec:Introduction}

Determining the nature of dark matter and detecting it has remained an outstanding problem in astrophysics. For several decades, significant evidence in favour of a cold dark matter universe has accumulated from observations of galactic clusters and large-scale structure (Davis et al. 1985), supernovae (Perlmutter et al. 1999) and the cosmic microwave background (CMB) anisotropies (Bennett et al. 2003).It is believed that if WIMPs populate the halo of the Milky Way today, then they could be detected via direct 
WIMP-nucleon interactions. Experiments for direct detection are currently underway, e.g.  PICASSO (Barnabé-Heider et al. 2005), CDSM (Akerib et al. 2006), SIMPLE (Alner et al. 2005), NAIAD (Girard et al. 2005), and DAMA (Bernabei et al. 2008). Unfortunately, to date, no such direct interactions have been observed. However, these direct detection experiments have been used for placing stringent limits on the WIMP-nucleon cross-section. 
These limits are obtained assuming a value of ${\rm \sim 0.3 \rm\ GeV/cm^3}$  for the local DM density in the Milky Way halo (Barnabé-Heider et al. 2005; Lewin $\&$ Smith 1996; Gates, Gyuk, $\&$ Turner 1995).

 WIMP-nucleon interactions in the Sun (Press $\&$ Spergel 1985; Silk, Olive, $\&$ Srednicki 1985) or in the Earth (Freese 1986; Gaisser, Steigman, $\&$ Tilav 1986; Krauss, Srednicki, $\&$ Wilczek 1986; Belotsky, Damour $\&$ Khlopov 2002) would cause the WIMPs to lose enough kinetic energy to become trapped in their 
 respective  potential wells, which could in turn create an observable neutrino flux from DM annihilation in the core. However, similar to direct detection experiments, no such neutrino signal has been observed. 

A crucial factor to consider in light of these null results is the irregularities in the distribution of DM in the halo.
 Density fluctuations on sub-parsec scales allow for the possibility for our solar system to reside in a region of significantly lower density than the average (Taylor $\&$ Babul 2005). If this is the case, then the actual WIMP-nucleon cross-section  could be higher than estimates which assume a local density equal to the average value of ${\rm \sim 0.3 \rm\ GeV/cm^3}$.  Furthermore, it could explain the lack of any direct detection of DM on Earth or any neutrino signal from the Sun. 

Extending this line of reasoning, we consider surveying the rest of the Galactic  disk for an expected neutrino signal from WIMP annihilations in other stars. A large-scale survey would alleviate any possibilities of lacking signals from DM voids due to regional density fluctuations. Estimates on high energy neutrino signals produced in various astrophysical processes has been a topic of interest for many years (for details, see Berezinsky $\&$ Zatsepin 1970; Berezinsky $\&$ Smirnov 1975; Stecker 1979). However, a new generation of neutrino telescopes is now opening a new window into the field. Large neutrino detectors (either built or planned) are capable of providing us with potentially ground-breaking observations in the fields of particle physics, astrophysics, and cosmology.
 Among the proposed detection techniques, under-ice and under-water optical Cerenkov detectors may prove to be particularly effective. This technique has already been implemented very successfully in the Lake Baikal experiment (Aynutdinov et al 2003), and in the AMANDA detector at the South Pole (Ackermann et al 2005). The much larger, ${\rm km^3}$-size detector IceCube (Ahrens et al. 2004) is currently under construction in the Antarctica, while other detectors (ANTARES, NEMO and NESTOR) are also being built in the Mediterranean Sea (Candia 2005).

In this paper, as we extend the analyses done on the solar capture and annihilation of WIMPs to all the stars in the Milky Way disk, we show that for any given line of sight along the disk, the estimated neutrino flux is larger than the expected flux from WIMP annihilations elsewhere in the Galaxy or halo. Identifying point sources may be difficult to achieve, since they are likely to be too weak for a detector to receive an unambiguous directional signal. Hence, we focus on detectible signals corresponding to unresolved diffuse fluxes.

In order to do this properly, a sound knowledge of other possible diffuse neutrino signals is also important, such as those produced from cosmic rays. It is well known that at low energies, CR interactions in the atmosphere produce mesons, which decay into a neutrinos (Volkova 1980; Gaisser $\&$ Honda 2002). To date, this is the only type of neutrino signal identified so far. Another contribution to the high energy diffuse neutrino fluxes is that resulting from the interactions of the CRs present all over the Galaxy with the ISM (Domokos, Elliott, $\&$ Kovesi-Domokos 1993; Berezinsky et al. 1993; Ingelman $\&$ Thunman 1996). The very low baryon density in the ISM ${\rm (\sim 1/cm^3 )}$ implies that essentially all mesons produced in such interactions decay without suffering any attenuation (Candia 2005). Measuring and comparing these fluxes could potentially give valuable information about the distribution and behavior of DM and cosmic rays in the galaxy.


\section{Dark Matter Annihilation \& Neutrino Sources}
\label{sec:dm}

Dark matter could exist in several forms. There have been several particle candidates proposed in the literature such as axions, neutralinos, and Kaluza-Klein particles, but speaking more generally here, we maintain the typical WIMP mass, $ M_{\chi}$, to be between a few GeV and a few TeV (for details see Jungman, Kamionkowski, $\&$ Griest 1996). In this work, we consider WIMPs in this mass range. If such a DM particle were in equilibrium with photons in the early universe, as the temperature decreased, a freeze-out would occur, leaving a thermal relic density. The temperature at which this occurs and the remaining DM density depends on the annihilation cross-section and mass of the WIMP. The cross-section is considered to be small enough to maintain a relatively slow and steady rate of dark matter annihilation even once large scale structure formation began and is still continuing within that structure today.

The rate of neutrino production in WIMP annihilations is highly model-dependent as the annihilation fractions to various products can vary largely from model to model. We do not consider direct WIMP annihilation chanels into neutrinos, as they are considered rare or nonexistent, but many indirect channels exist. WIMPs typically annihilate into pairs of quarks or vector bosons. The quarks may then fragment into hadronic jets containing a large number of particles. The mass of the parent DM particle will effect what kinds of decay chains will follow from the annihilations. However, by setting the mass of the WIMP greater than the top quark, $M_\chi>M_{\rm t}$, where $M_{\rm t}= 175 \rm\ GeV$, calculating the final neutrino outcome is greatly simplified. The reason is that the top quark lifetime is short, and almost $100\%$ of the time decays as $t \rightarrow W $. The $W$ boson, like the $t$, decays before virtually any energy loss and creates neutrinos with equal branching ratios (each about 10.5\%) into $e\nu_{e}$, $\mu\nu_{\mu}$, and $\tau\nu_{\tau}$ (Crotty 2002). Therefore, this is the only decay chain that will need to be considered.

The two main sources of neutrinos from DM annihilation in the galaxy that we are considering here are from the general annihilations occurring in the halo, and for WIMPs that have been captured and have annihilated in the stars. Signals from general DM annihilation throughout the Galactic halo have been considered by (Finkbeiner 2005, Baltz $\&$ Wai 2004), yet an enhanced signal ought to be present from DM annihilations inside the stars. In general, dark matter particles must be gathered in high concentrations to provide an observable flux of neutrinos. The potential wells of the Sun, Earth, stars or galactic centre are examples of regions where such concentrations may occur. Gravitational interactions bring a portion of the WIMPs in the Galactic halo and disk into the stars. For main sequence stars, we assume that the star is in hydrostatic equilibrium and is spherically symmetric, and that effects of rotation and magnetic fields are negligible (Lopes, Bertone, $\&$ Silk 2002). When WIMPs enter a star, they may interact with nuclei and lose enough kinetic energy to be trapped by the gravitational potential well and eventually annihilate with each other. 

There are two different channels by which WIMPs can scatter off nuclei in a given star: spin-dependent (axial) interactions and spin-independent (scalar) interactions. The spin-dependent interactions refer to WIMP-on-proton interactions, while spin-independent interactions refer to WIMP interactions with the heavier elements in the star. Since an average star's content consists primarily of protons, the spin-dependent cross-section is typically several orders of magnitude larger than the spin-independent cross-section, and therefore, spin-dependent scattering dominates (Hooper $\&$ Silk 2004). 

As the WIMPs annihilate in the star's core, most of their decay products are subject to absorption and/or diffusion, but the neutrinos stand the best chance of escaping the star and propagating to earth relatively undisturbed. For a maximal neutrino signal, we assume at present that the WIMP capture and annihilation rates are in equilibrium, $\Gamma_{\rm A} = \Gamma_{\rm C}/2$, (each annihilation destroys two WIMPs). Among the varieties of neutrinos produced, the muon neutrinos, $\nu_{\mu}$, are particularly useful for indirect detection of DM annihilation processes (hereafter 'neutrinos' specifically implies muon neutrinos). When the neutrinos collide with earth materials they produce muons, which have a relatively long range in a detector medium like ice or water. For more details on this detecting process of galactic neutrinos, see the results of (Abe et al. 2006; Kelly et al. 2005).


\section {Calculating the Neutrino Spectra from WIMP Annihilation in Stars}
\label{sec:Dark Matter Candidates}

 Assuming the Galaxy to be rotationally symmetric in the plane, the neutrino flux produced from WIMPs that have been captured by and annihilated in stars, $\Phi_{\nu_{*}}$, can be determined from the general expression

\begin{equation}
\Phi_{\nu_*}(l) =\frac{1}{4 \pi}\int_{r_{1_{\rm cut}}}^{r_{1_{\rm max}}} {\epsilon(r_1)  \rm\ \dd r_1}.
\end{equation}
Here $r_{1_{\rm cut}} < $$r_1$$ < r_{1_{\rm max}}$, where $r_1$ represents the distance along a line of sight (LOS), $r_{1_{\rm cut}}$ is the cut-off distance from the Earth, which we take to be $\sim~1000$ AU and $r_{1_{\rm max}}$ is a great distance where the number density of stars in the disk is zero. We define $r$ as the distance from the Galactic center to the source of the annihilation, geometrically related to $r_1$ as

\begin{equation}
r(l) = \sqrt{r_\odot^2 + r_1^2 - 2\cos(l) r_1 r_\odot}
\end{equation}
where $l$ is the angle between the LOS to the Galactic center and the LOS to the source of the annihilation ($r_1$) for an observer on Earth. The emissivity at  a point $r_1$ is related to the dark matter density, $ \rho_{\rm DM}$, 
the stellar mass density,  $\rho_*(r)$, and 
the emission rate per solar size star, $L_\nu$, and is as follows, 

\begin{equation}
\epsilon(r_1)=\biggl[\frac{\rho_{\rm DM}(r)}{\rho_\odot}\biggr]\rho_*(r) L_\nu\; .
\end{equation}
In this expression $\rho_\odot = 6.5\times 10^{-25} {\rm g/cm^3}$ is the dark matter density in our local region of the galaxy and $\rho_{\rm DM}(r)$ is the dark matter density at a point $r$, which we approximate with a Naverro-Frenk-White profile (Navarro, Frenk, $\&$ White 1996)
\begin{equation}
\rho_{\rm DM}(r) = \frac{\rho_or_{\rm s}^3}{r(r^2 + r_{\rm s}^2)}.
\end{equation}
where $\rho_o = 5.6\times 10^{-25}\;  \rm g/cm^3$, and the scale radius, $\rs = 20 \rm\ kpc$, (Finkbeiner 2005). The number density of stars in the midplane of the disk was estimated according to the following expression from (Efstathiou 2000) where

\begin{equation}
\rho(z)=\frac{\mu_{\rm d}}{2\hd}{\rm sech}^2\Big(\frac{z}{\hd}\Big).
\end{equation}

Here $\hd $ is the thickness of the galactic disk, $200 \rm\ pc$, and $\mu_{\rm d}=\mu_o e^{-r/3.5{\rm kpc}}$ is the surface mass density of stars at a distance $r$ from the galactic center. Considering lines of sights in the midplane,  $z=0$, we get 
\begin{equation}
\rho_*(r) = \frac{\mu_{\rm o} e^{-r/3.5{\rm kpc}}}{ 2 \hd}
\end{equation}
where $\mu_{\rm o} = 567 M_\odot/{\rm pc}^2$.   In calculating the emission rate, $L_{\nu}$, we account for all neutrinos within the energy range, $E_{\rm min} < E_{\nu} < M_{\chi}$, where $E_{\rm min}$ is set by a typical detector threshold of 50 GeV (Crotty 2002). 
\begin{equation}
L_{\nu} = \frac{dN}{dt} = \int_{E_{\rm min}}^{M_{\chi}}\frac{dN}{dEdt} \dd E
\end{equation}
where the emergent flux is
\begin{equation}
\frac{dN}{dEdt} = \Gamma_{\rm A} \frac{dN}{dE} P.
\end{equation}
Here $\Gamma_{\rm A}$ is the annihilation rate of the WIMPs inside the star, $\frac{dN}{dE}$ is the initial energy distribution of the neutrinos produced per annihilation, and $P$ is the probability that a neutrino produced in the star's core will emerge from its edge without being absorbed by the steller material. Assuming that all the WIMPs annihilate at a rate equal to the rate at which they are captured, we define the annihilation rate as the product of the DM flux and the WIMP-nucleon cross-section: 

\begin{equation}
\Gamma_{\rm A} = \phi_{\chi} \sigma_{*}
\end{equation}
The DM flux, $\phi_{\chi}$, is defined as (Barger 2002)
\begin{equation}
\phi_{\chi} = \frac{1.2 \times 10^7}{M_{\chi}} \rm\ cm^{-2} s^{-1}.
\end{equation}
Assuming that the sun is an average size star, we may assume a reasonable capture cross-section for the sun $\sigma_\odot$ to be similar to the average capture cross-section for the stars, $\sigma_{*}$, in the galactic disk. 

\begin{equation}
\sigma_{*} = \sigma_\odot = f(1.2 \times 10^{57})\sigma_{\rm el}
\end{equation}
where $f$ = 10 is the focusing factor, i.e. the ratio of kinetic and potential energy of the WIMP near a given star, and $1.2 \times 10^{57}$ is the number of nucleons in stars the size of our sun.  The choice of parameters needed to compute WIMP-nucleon elastic scattering, $\sigma_{\rm el}$, is at best approximate and are difficult to obtain with any generality as terrestrial experiments have indicated (Barnabé-Heider et al. 2005). To this extent, we rely on the results of the various experiments, (Barnabé-Heider et al. 2005, Gates, Gyuk, $\&$ Turner 1995, Lee et al. 2007) and take the most conservative cross-section estimates of WIMP-proton and WIMP-neutron interactions. We use a weighted-sum of the two, working out to ${\sim0.5 - 1\rm pb}$ which may vary slightly with WIMP mass. However, accounting for small scale  DM density fluctuations, the WIMP-nucleon cross-section could be significantly higher than what we use here.

The spectrum of neutrinos, $\frac{dN}{dE}$, is straightforward to calculate. Resulting from the decay chain $t \rightarrow W \rightarrow \nu$, we definine $x$ as $E/M_\chi$ and integrate over the fragmentation function of (Albuquerque, Hui, $\&$ Kolb 2001) to first compute the total number of hadrons, $N_H$, produced per annihilation.

\begin{equation}
N_{\rm H} = \int_{1}^{\eta} \frac{dN_{\rm H}}{dx}\dd x
\end{equation}
where $\eta = \Lambda_{\rm QCD}/M_{\chi}$, with the QCD scale set as $\Lambda_{\rm QCD} = 0.1 \rm\ GeV$, and the fragmentation function is 

\begin{equation}
\frac{dN_{\rm H}}{dx} = a x^{-3/2}(1 - x)^2,
\end{equation}
$a = 15/16$. The number of top quarks that are produced per annihilation is determined simply by the expression

\begin{equation}
N_{\rm t} = N_{\rm H} \sqrt{\frac{\Lambda_{\rm QCD}}{M_{\rm t}}}.
\end{equation}
From the number of top quarks produced per annihilation we can estimate the number of muon neutrinos that follow by assuming a branching fraction, $B$, of $\sim 10\%$ (Albuquerque, Hui, $\&$ Kolb 2001; Barger 2002). The number of top quarks and corresponding neutrinos produced per annihilation varies with the mass of the WIMP as indicated in Figure 1.

The initial energy spectrum of the neutrinos produced per annihilation in the stellar core is: 

\begin{equation}
\frac{dN}{dE} =  \frac{B N (1-x)^2(E + M_{\rm W})}{\sqrt{(E + M_{\rm t})((E + M_{\rm t})^2 - M_{\rm t}^2) ((E+ M_{\rm W})^2 - M_{\rm W}^2)}}. 
\end{equation}

Here $N = 100$ is the normalization factor, $B$ is the branching ratio,  $x = E/M_{\chi}$ is the fraction of the initial WIMP energy given to the neutrino, and $M_{\rm W} = 80 \rm\ GeV$ is the mass of the W boson. We also account for the possibility that when the $\tau\nu_{\tau}$ pair is produced by the W decay, the $\tau$ also decays before losing much energy, producing a secondary $\nu_{\mu}$ around $18\%$ of the time, (Crotty 2002).

The survival probability of these neutrinos must be considered. Since a star is opaque to energetic neutrinos, the emergent energy spectrum per annihilation is not the same as the core spectrum. Some neutrinos which are produced in a stellar core suffer charged-current (CC) interactions and are rapidly thermalized as they make their way to the edge to escape. CC interactions remove  a fraction of neutrinos above a certain transparency energy $E_{\rm k}$, where $\sigma(E_{\rm k})\int n(r) dr$ becomes unity (Ritz $\&$ Seckel 1988). Thus, the probability of a neutrino escaping a star without interaction is given by

\begin{equation}
P = e^{-E/E_{\rm k}}
\end{equation}
where $E_{\rm k} = 150 \rm\ GeV$ (Albuquerque, Hui, $\&$ Kolb 2001). A comparison of these spectra of neutrinos produced in the core and surviving to the edge of a star for different WIMP masses can be seen in figure 1.

\begin{figure}
\centering
\mbox{\epsfig{figure=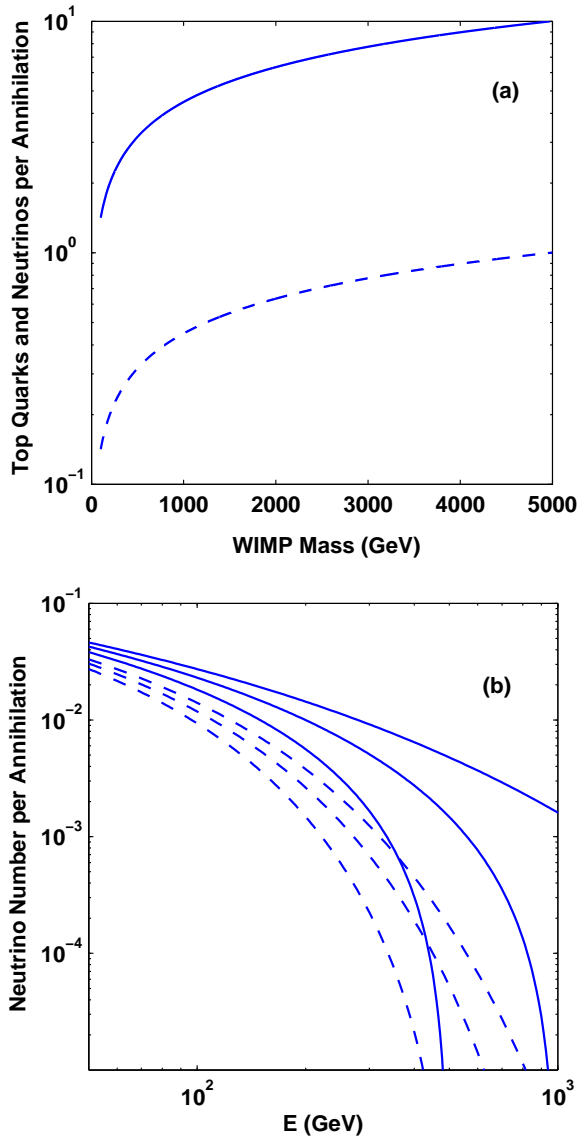,height=6.2in}}
\vspace{-0.6cm}
\caption{(a) Number of top quarks (solid line) and neutrinos (dashed line) produced per WIMP annihilation as a function of WIMP mass. (b) The energy distribution of neutrinos produced per annihilation inside the stellar core (solid lines) and the final spectrum emitted from the stellar edge (dashed lines) for WIMP masses of (top to bottom) 500 GeV, 1000 GeV, and 5000 GeV, respectively.}
\label{fig:rates}
\end{figure}


\section {Neutrino Spectra from Dark Matter Annihilation in Halo}
\label{sec:Dark Matter Candidates}

Using the same general expression as equation (1) and integrating over the same range, the neutrino flux from the 'regular' DM annihilations in the disk and halo is calculated by setting

\begin{equation}
  \epsilon(r_1) = \Big[\frac{\rho_{\rm DM}(r)}{\rho_\odot}\Big]^{-2} B \Big(\frac{\rho_\odot}{m_\chi}\Big)^2 \langle\sigma v\rangle 
\end{equation}
where $m_\chi$ is the mass of a WIMP in grams and $\langle\sigma v\rangle$ =  $2 \times 10^{-26} {\rm cm^3/s}$ is the DM annihilation cross-section, (as appears in Finkbeiner 2005).


\section {Neutrino Spectra from Cosmic Ray Interactions with ISM}

According to the standard picture, Galactic CRs are accelerated by the expanding shock wave fronts produced by the explosions of supernovae. This mechanism is capable of producing cosmic ray power spectrum with an index of $\sim 2.5 - 3$ (Strong, Moskalenko, $\&$ Ptuskin 2007). As cosmic rays propagate through the Galaxy, they interact with the ISM, which is a very low density, non-relativistic plasma constituted mainly by atomic and molecular hydrogen. The mesons produced in these proton-proton interactions decay before losing energy in secondary interactions. The muons produced from the meson decay then decay themselves and produce neutrinos. 

Given the uncertainties concerning the matter distribution in the Galaxy, one can only make a fairly simple model for it. The model used here is similar to previous ones (Domokos, Elliott, $\&$ Kovesi-Domokos 1993; Berezinsky et al. 1993; Ingelman $\&$ Thunman 1996; Candia 2005), relatively standard, and should be adequate for our purpose of estimating the neutrino flux from CR-ISM interactions. 

In this short study of neutrino fluxes from CRs, interactions with stars are not taken into account. Even though CRs can indeed interact with the stellar material and produce mesons, neutrinos and gamma-rays, these by-products must traverse the star such that the photon and neutrino fluxes would be significantly attenuated to a negligible level. 

In this scenario, in order to make a comparison with the neutrinos produced from WIMP annihilations we integrate over the same energy range here as well, $E_{\rm min} < E < M_\chi$. The full expressions for calculating the neutrino flux can be found in (Candia 2005). Here we present the abbreviated expressions, i.e. the resulting emissivity in a parameterised form inspired by the analytic formalism of (Berezinsky et al. 1993; Ingelman $\&$ Thunman 1996). Using the same general expression as equation (1) the neutrino flux from CR-ISM interactions in the disk is obtained with

\begin{equation}
\epsilon(r_1)= \int_{E_{\rm min}}^{E_{\rm max}} \delta(r) N_{\rm o} E^{-\gamma - 1}\dd E
\end{equation}
where $N_{\rm o} = 3 \times 10^{-6}$, $\gamma = 1.63$ is the spectral energy index, consistent with standard estimates for cosmic ray power-law spectra, (Berezinsky et al. 1993; Strong, Moskalenko, $\&$ Ptuskin 2007). We smooth over the inhomogeneities of the ISM density with a modest exponential function decreasing by a factor of  $\sim 2$ from Galactic center to edge; ranging from ${\rm \sim 2 \rm\ nucleons/cm^3}$ to ${\rm \sim 1 \rm\ nucleon/cm^3}$, respectively. 

\begin{equation}
\delta(r) = \delta_{\rm o} e^{-r/5{\rm kpc}}
\end{equation}
where the values of $\delta_{\rm o} = 2.2$ and $5 \rm\ kpc$ were chosen to fit with the data provided by (Berezinsky et al. 1993).
The interactions between CR particles and the ISM produce more $\mu$ neutrinos than $e^-$ neutrinos, however, by the time they arrive to Earth the neutrinos flavors are evenly distributed due to vacuum oscillations during propagation (Athar, Jezabek, $\&$ Yasuda 2000). Since, the ISM and the CR densities are largest in the Galactic center region, there is a slight but noticeable anisotropy, being maximal in the direction to the Galactic center while being lower in the orthogonal planar direction as can be seen in figure 2 (thick solid lines). The comparison of the fluxes are presented in the next section.


\section {Results}
In figure 2 the neutrino fluxes from WIMP annihilations and CR-ISM interactions are presented as a function of galactic longitude, $l$, for different WIMP masses and minimal energy thresholds. Figure 2(a) compares neutrino fluxes from general DM annihilations, enhanced annihilations in stars, and from CR-ISM interactions for different values of $M_\chi$. These spectra were all integrated from $E_{\rm min} = 50 \rm\ GeV$ up to a maximum energy of $M_\chi$. For each of the given WIMP masses, the neutrino signal from WIMP annihilations in stars is higher than general WIMP annihilations in the halo by at least an order of magnitude. For values of $M_\chi = 1000 \rm\ GeV$ or lower, the neutrino signal from WIMP annihilation in stars is above the flux from CR-ISM interactions for $\sim 5^{\circ}$ or $10^{\circ}$ from the galactic center. The flux produced from annihilations of heavier WIMPs would fall below the CR-ISM signal, primarily due to the significant attenuation the neutrinos suffer as they try to escape from the stars.   

Figure 2(b) compares neutrino fluxes from general DM annihilations, enhanced annihilations in stars, and from CR-ISM interactions for different minimal energy thresholds. These spectra were all integrated up to a maximum energy of $M_\chi = 1000 \rm\ GeV$. Again, the neutrino signal from WIMP annihilations in stars remains consistently higher than general WIMP annihilations in the halo. For values of $E_{\rm min} = 250 \rm\ GeV$ or lower (well within detector constraints), the neutrino signal from WIMP annihilations in stars is above the flux from CR-ISM interactions for $\sim 5^{\circ}$ or $10^{\circ}$ from the galactic center. By $E_{\rm min} = 500 \rm\ GeV$ the signal from CR-ISM dominates over the signal from WIMP annihilation.

Fixing the viewing angle to be looking at the galactic center, Figure 3 compares neutrino fluxes from general DM annihilations in the halo, enhanced annihilations in stars, and from CR-ISM interactions as a function of energy for different values of $M_\chi$. Similar to Figure 2(a), these spectra were all integrated from $E_{\rm min} = 50 \rm\ GeV$ up to a maximum energy of $M_\chi$. In this figure, the neutrino signal from WIMP annihilations in stars is higher than general WIMP annihilations in the halo by at least a couple orders of magnitude. For designated values of $M_\chi = 500 \rm\ GeV$ or $M_\chi = 1000 \rm\ GeV$, the neutrino signal from WIMP annihilation in stars is above the flux from CR-ISM interactions for energies up to several hundred GeV. These conditions of an enhanced WIMP signal above the diffuse signal from CR-ISM interactions ought to be clearly observable within a fixed viewing angle and energy range.

\begin{figure}
\centering
\mbox{\epsfig{figure=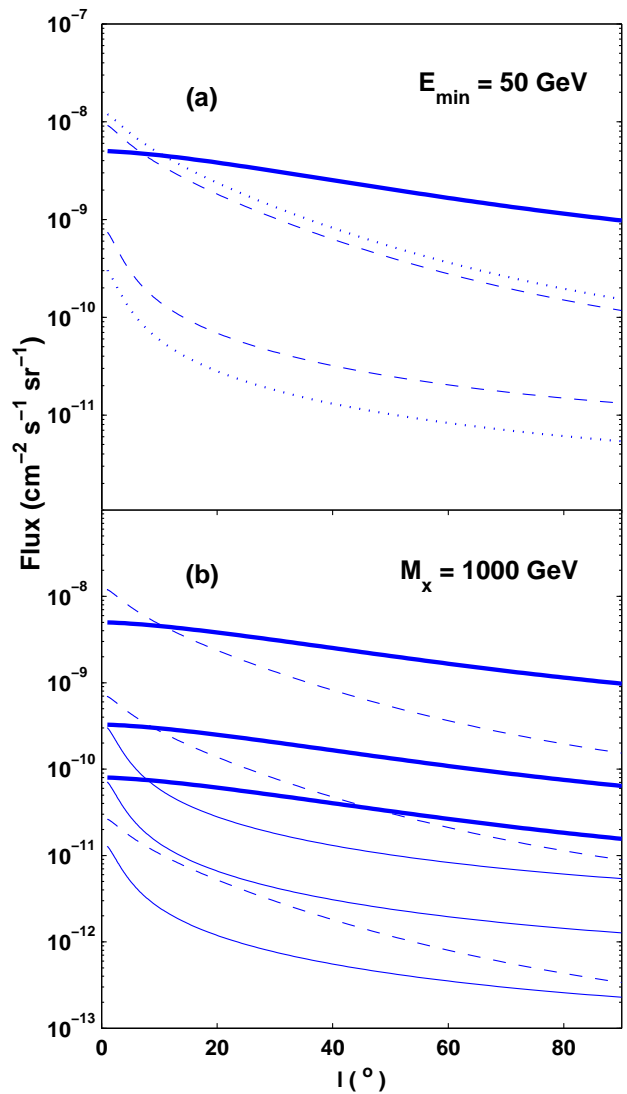,height=6in}}
\vspace{-0.6cm}
\caption{(a) Neutrino fluxes from CR-ISM interactions (thick solid line), general DM annihilations, and enhanced annihilations in stars as a function of galactic longitude for different values of $M_\chi$. For $M_\chi = 500 \rm\ GeV$ (dashed lines) and $M_\chi = 1000 \rm\ GeV$ (dotted lines), the bottom and top curves represent the general and enhanced neutrino signals, respectively. These spectra were all integrated from $E_{\rm min} = 50 \rm\ GeV$ up to a maximum energy of $M_\chi$.  (b) Similar neutrino fluxes as 2(a) but for different minimal energy thresholds. The thin solid and dashed lines represent the general and enhanced neutrino signals, respectively. These spectra were all integrated up to a maximum energy of $M_\chi = 1000 \rm\ GeV$. Read from top to bottom, all the lines represent the neutrino spectra with $E_{\rm min}$: 50 GeV, 250 GeV, and 500 GeV, respectively.}
\label{fig:rates}
\end{figure}

\begin{figure}
\centering
\mbox{\epsfig{figure=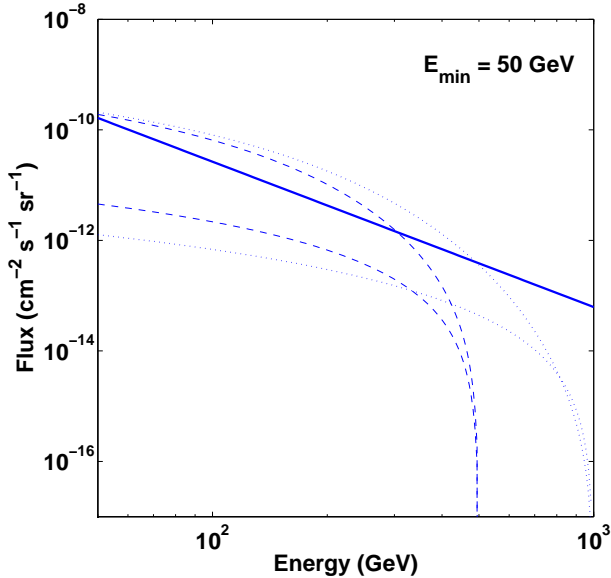,height=3.2in}}
\vspace{-0.6cm}
\caption{Looking toward the galactic center, neutrino fluxes from CR-ISM interactions (thick solid line), general DM annihilations, and enhanced annihilations in stars as a function of energy for different values of $M_\chi$. For $M_\chi = 500 \rm\ GeV$ (dashed lines) and $M_\chi = 1000 \rm\ GeV$ (dotted lines), the bottom and top curves represent the general and enhanced neutrino signals, respectively. These spectra were all integrated from $E_{\rm min} = 50 \rm\ GeV$ up to a maximum energy of $M_\chi$.}
\label{fig:rates}
\end{figure}


\section{Discussion \&\ Conclusions}\label{sec:conclusions}

Detection of WIMP annihilation into neutrinos is a potentially promising indirect detection method. A small fraction of the neutrinos that reach the detector are converted into muons through charged current interactions (Gandhi et al. 1996). These muons then propagate through the Cerenkov medium of the detector, where they are detected by photomultiplier tubes distributed through the effective volume. For a review of high-energy neutrino astronomy, see (Halzen $\&$ Hooper 2002). As a muon propagates, it loses energy at a given rate and then may eventually fall below the energy threshold of the detector, typically 10-100 GeV for deep ice or water detectors. $E_{\mu}$ can be estimated as 
$\sim E_{\nu}/2$. For experiments with energy thresholds like these, lighter WIMPs can be very difficult or impossible to detect for this reason. The neutrino-induced muon flux may be detected in a neutrino telescope by measuring the muons that come from stars in the direction of a line of sight in the Galactic disk. These neutrinos should have energies well above the energy of solar neutrinos and so should be easily distinguishable. 

At present, the AMANDA experiment is the largest high-energy neutrino telescope online, with an effective
area of 50 000 ${\rm m^2}$ and a muon energy threshold of about 30 GeV. It is currently taking data at the South Pole (Andres et al. 2001), and soon to be superseded by the much larger IceCube (Ahrens et al. 2004), under construction there. IceCube will have a somewhat higher energy threshold (50 - 100 GeV), but a full square kilometre of effective area. Also under construction, at higher latitudes in the Mediterranean, ANTARES (Blanc et al. 2003) will be significantly more sensitive to dark matter annihilations compared to AMANDA and IceCube due to its lower energy threshold (10 GeV) and larger effective area than AMANDA (100 000 ${\rm m^2}$). And unlike the south pole telescopes, ANTARES will also be sensitive to neutrinos coming from the Galactic centre. These detector thresholds are well within the desginated WIMP masses chosen in this work.

These kinds of experiments will have background events consisting of atmospheric neutrinos (Gaisser, Halzen, $\&$ Stanev 1995). In a given direction (up to the angular resolution of a neutrino telescope), tens of events $>$100 GeV ${\rm km^{-{2}} yr^{-1}}$ are expected from the atmospheric neutrino flux. Fortunately, for detectors with very large-volumes  and sufficient statistics, it should be possible to significantly reduce this background or eliminate it entirely. 

Given the current limits of AMANDA, detecting a signal from WIMP annihilations is not very strong. However, experiments with lower energy thresholds (ANTARES) and larger effective areas (IceCube) will greatly enhance the sensitivity and chances for detection in coming years.

While this indirect detection method may appear promising in some respects, it must also be kept in mind that the rate of neutrinos produced in WIMP annihilations is highly model-dependent. The sensitivity of a square-kilometre neutrino detector with a moderate muon energy threshold (50 GeV) will yield an observable signature in only a fraction of possible DM models, although such experiments are nevertheless an important probe. Accurately modeling the capture and annihilation rate of WIMPs by stars depends critically on DM distribution in the halo, and the WIMP-nucleon cross-section, which are all central to the prediction of the neutrino flux (and therefore muon flux) that is expected to be measured by neutrino telescopes.

While most of our understanding of the properties of DM on subgalactic scales have come from numerical simulations of structure formation (De Lucia et al. 2004; Diemand, Moore, $\&$ Stadel 2004; Gao et al. 2004), they currently lack the resolution necessary to accurately estimate the level of substructure on the subparsec level. In which case, without such resolving power, it remains a definite possibility  that the DM density in our local solar region is lower than the average DM density at the solar distance due to density fluctuations. We have thus chosen a very conservative value for the WIMP-nucleon cross-section based on estimates from terrestrial direct-detection experiments. Such a conservative cross-section would produce a rather modest capture and annihilation rate in stars. However, by surveying the stellar disk, we may see that the output flux from DM annihilations in distant stars could actually be higher than what is emitted from the Sun.  

The neutrino flux arising from CR-ISM interactions can be easily distinguished from the flux from DM by extending the neutrino survey up to higher energies. Given the high energies that cosmic rays can get to, they are capable of producing much higher energies than could be produced from DM annihilations, given the upper limits on the estimated WIMP mass. Also, at the relevant energy levels, the neutrinos produced from the CRs should be accompanied by a photon flux (Berezinsky et al. 1993), which would not be present from the DM annihilations occurring in stars. A diffuse gamma-ray flux above TeV energies from CR-ISM interactions could be measured with such ground-based detectors as CANGAROO, VERITAS, H.E.S.S. and Milagro, (Candia 2005). In fact, the Milagro Gamma Ray Observatory has already reported the first observations of a diffuse  gamma-ray signal at TeV energies in the galactic plane (Atkins et al 2005). Besides the pion decay channel, diffuse gamma-rays can also be produced by high energy electron bremstrahlung and inverse Compton scattering with the interstellar radiation field. For a detailed calculation of the photon flux produced from  CR-ISM interactions, see (Candia 2005).

Being able to escape from very dense astrophysical environments and travel vast distances undeflected and unattenuated, neutrinos offer precious information about DM as well as the high energy Universe. Such signatures will surely be subject to extensive experimental investigations in view of the new neutrino telescopes planned or under construction.


\section*{Acknowledgements}
This research is supported by the German-Israeli Science Foundation for Development 
and Research and by the Asher Space Research fund.
ZM wishes to thank the Israeli Ministry of the Absorption of Science for providing the research funds necessary to support this work.


\end{document}